\documentclass[useAMS,usenatbib]{mnras}
\usepackage[dvips]{graphicx}
\usepackage[english]{babel}
\usepackage{amsmath}
\usepackage{amssymb}
\usepackage{textcomp}
\usepackage{natbib}

\title[SZ effect and soft X-rays from galaxy haloes]{The impact of stellar feedback on hot gas in galaxy haloes: the Sunyaev-Zel'dovich effect and soft X-ray emission}
\author[F. van de Voort et al.]{Freeke~van~de~Voort$^{1,2}$\thanks{E-mail: freeke@berkeley.edu}, 
Eliot~Quataert$^{1}$,
Philip~F.~Hopkins$^{3}$, 
\newauthor
Claude-Andr\'e~Faucher-Gigu\`ere$^{4}$,
Robert~Feldmann$^{1}$,
Du\v{s}an~Kere\v{s}$^{5}$,
T.~K.~Chan$^5$
\newauthor
and Zachary~Hafen$^4$ \\
$^{1}$Department of Astronomy and Theoretical Astrophysics Center, University of California, Berkeley, CA 94720-3411, USA \\
$^{2}$Academia Sinica Institute of Astronomy and Astrophysics, P.O. Box 23-141, Taipei 10617, Taiwan \\
$^{3}$TAPIR, Mailcode 350-17, California Institute of Technology, Pasadena, CA 91125, USA \\
$^{4}$Department of Physics and Astronomy and CIERA, Northwestern University, 2145 Sheridan Road, Evanston, IL 60208, USA \\
$^{5}$Department of Physics, Center for Astrophysics and Space Science, University of California at San Diego, 9500 Gilman Drive, \\
\ \ La Jolla, CA 92093 
}

\begin{document}

\date{Accepted 2016 September 11. Received 2016 August 16; in original form 2016 April 05}

\pagerange{\pageref{firstpage}--\pageref{lastpage}} \pubyear{2016}

\maketitle

\label{firstpage}

\begin{abstract}

The thermal Sunyaev-Zel'dovich (SZ) effect and soft X-ray emission are routinely observed around massive galaxies and in galaxy groups and clusters. We study these observational diagnostics of galaxy haloes for a suite of cosmological `zoom-in' simulations from the `Feedback In Realistic Environments' project, which spans a large range in halo mass ($10^{10-13}$~M$_\odot$). We explore the effect of stellar feedback on the hot gas observables. The properties of our simulated groups, such as baryon fractions, SZ flux, and X-ray luminosities ($L_X$), are broadly consistent with existing observations, even though feedback from active galactic nuclei is not included. We make predictions for future observations of lower-mass objects for both SZ and diffuse X-ray measurements, finding that they are not just scaled-down versions of massive galaxies, but more strongly affected by galactic winds driven by star formation. Low-mass haloes ($\lesssim10^{11}$~M$_\odot$) retain a low fraction of their baryons, which results in a strong suppression of the SZ signal. Our simulations therefore predict a scaling with halo mass that is steeper than self-similar for haloes less massive than $10^{13}$~M$_\odot$. For halo masses $\lesssim10^{12}$~M$_\odot$, $L_X$ is time-variable and correlated primarily with the star formation rate (SFR). For these objects, the diffuse X-ray emission is powered mostly by galactic winds and the gas dominating the X-ray emission is flowing out with radial velocities close to the halo's circular velocity. For halo masses $\gtrsim10^{13}$~M$_\odot$, on the other hand, $L_X$ is much less variable and not correlated with the SFR, because the emission originates from the quasi-hydrostatic, virialized halo gas.

\end{abstract}


\section{Introduction}

Hot gaseous haloes around galaxies are an integral part of the galaxy formation process. As matter in the Universe collapses, it forms a network of sheets and filaments, the so-called `cosmic web', along with nearly-spherical dark matter haloes. The collapse of dark matter halts as it reaches virial equilibrium in haloes. Baryons, on the other hand, can lose energy through radiation, which allows them to collapse further and reach high densities in the centre, where they settle in a rotationally supported disc \citep[e.g.][]{Fall1980}. According to the simplest picture of baryonic collapse, all gas in a dark matter halo is initially heated to the virial temperature of that halo at a virial shock, within which it reaches a quasi-static equilibrium supported by the pressure of the hot gas. 

However, within the so-called cooling radius, the cooling time of the gas is shorter than the age of the Universe. If the cooling radius lies well inside the halo, which is indeed the case for high-mass haloes, a quasi-hydrostatic, hot atmosphere will form. On the other hand, if the cooling radius is larger than the virial radius, as is the case for low-mass haloes ($<10^{12}$~M$_\odot$), there will be no hot halo and the accreting gas will not go through an accretion shock at the virial radius. Instead, the gas can cool and reach the central galaxy rapidly \citep{Rees1977, White1978, White1991, Birnboim2003}. The filamentary and clumpy structure of the intergalactic medium can give rise to hybrid haloes where both gas accretion modes coexist \citep[e.g.][]{Keres2005}. 

In reality the properties of gaseous haloes are more complicated than in this simple picture. Besides star formation acting as a sink for dense, cold gas, massive stars also pump energy into the halo through galactic winds, changing the structure and thermodynamics of the halo gas. Low-mass galaxies may lack a hot halo due to structure formation, but they may still be surrounded by hot gas due to galactic outflows and the interaction of such outflows with inflowing gas. Besides pumping energy into the halo, galactic outflows also remove gas from the halo, thus decreasing the amount of hot halo gas. This has strong observational consequences, which are detectable with current and future instrumentation. Understanding the detailed effect of stellar feedback on halo gas is one of the keys to understanding how galaxies are fed and how they are quenched.

Two main methods are employed to detect the hot gas around galaxies. The thermal Sunyaev-Zel'dovich (SZ) effect \citep{Sunyaev1970} is the increase in the energy of the cosmic microwave background (CMB) radiation caused by inverse Compton scattering of CMB photons off free electrons in hot gas. It thus depends linearly on the electron column density along the line-of-sight as well as on the temperature of the gas. Large samples of SZ observations have been and are being obtained with e.g.\ the Atacama Cosmology Telescope \citep{Menanteau2010}, the South Pole Telescope \citep{Vanderlinde2010}, and the Planck survey \citep{Planck2011, Planck2013}. 

Gas with temperatures $T\gtrsim10^6$~K also emits continuum and line radiation in the soft X-ray regime (0.5 to several~keV). This emission depends on the square of the density and on metallicity and temperature. Satellites, such as ROSAT, Chandra, and XMM-Newton, have made it possible to observe the extended hot halo gas in clusters and groups \citep[e.g.][]{Voges1999, Sun2009, Anderson2015}, around quiescent, elliptical galaxies \citep{Forman1985, OSullivan2001, Boroson2011, Goulding2016}, and around star-forming, spiral galaxies \citep[e.g.][]{Strickland2004, Tullmann2006, Anderson2011, Dai2012, Bogdan2013}. The latter generally show a correlation between X-ray emssion and star formation rate \citep[SFR; e.g.][]{Mineo2012, Li2013b}. This emission is necessarily biased towards high-temperature, high-metallicity, and high-density gas and therefore strongly biased towards the central regions of the halo. SZ and X-ray measurements thus probe different halo gas and combining these observations with cosmological simulations will allow us to untangle the complicated interplay between halo gas and galactic winds.

In this paper, we present results from a suite of cosmological `zoom-in' simulations from the `Feedback In Realistic Environments' (FIRE) project\footnote{http://fire.northwestern.edu/}, which spans a large range in halo mass. The FIRE simulation suite has been shown to successfully reproduce a variety of observations. This includes, e.g., the derived stellar-to-halo mass relationship \citep{Hopkins2014FIRE, Feldmann2016}, shallow dark matter profiles \citep{Onorbe2015, Chan2015}, the mass-metallicity relation \citep{Ma2016}, and high redshift H\,\textsc{i} covering fractions \citep{Faucher2015, Faucher2016}. The hot gaseous haloes around galaxies are the focus of this paper. We compare our simulated galaxy haloes to observations of the thermal SZ effect and to soft X-ray observations.
We study the properties of halo gas for halo masses below and above that of the Milky Way in order to determine in which regime stellar feedback significantly affects hot gas observables by changing the density, temperature, pressure, and metallicity of the hot halo gas and by removing gas from the halo. 

In Section~\ref{sec:sim} we describe the suite of simulations used, as well as the way we compute the thermal SZ effect (\ref{sec:SZmethod}) and soft X-ray emission (\ref{sec:Xmethod}) from the gas. In Section~\ref{sec:results} we present our results, with Section~\ref{sec:SZ} focusing on the thermal SZ effect and Section~\ref{sec:Xray} focusing on the soft X-ray emission from the hot haloes. The main results are shown in Figures~\ref{fig:Y_mass}, \ref{fig:X_mass}, and~\ref{fig:X_SFR}. We discuss and compare our results to those in the literature and conclude in Section~\ref{sec:concl}.

\section{Method} \label{sec:sim}

The simulations used are run with \textsc{gizmo}\footnote{http://www.tapir.caltech.edu/phopkins/Site/GIZMO.html} \citep{Hopkins2015} in `P-SPH' mode, which adopts the Lagrangian `pressure-entropy' formulation of the smoothed particle hydrodynamics (SPH) equations \citep{Hopkins2013PSPH}. The gravity solver is a heavily modified version of \textsc{gadget}-2 \citep{Springel2005}, with adaptive gravitational softening following \citep{Price2007}. Our implementation of P-SPH also includes substantial improvements in the artificial viscosity, entropy diffusion, adaptive timestepping, smoothing kernel, and gravitational softening algorithm. 

This work is part of the FIRE project, which consists of several cosmological `zoom-in' simulations of galaxies with a wide range of masses, simulated down to $z=0$ (\citealt{Hopkins2014FIRE, Chan2015, Ma2016, Hafen2016}; Feldmann et al.\ in prep.) and $z=2$ \citep{Faucher2015} and $z=1.7$ \citep{Feldmann2016}. The simulation details are fully described in \citet{Hopkins2014FIRE} and references therein. 

\begin{table*}
\begin{center}
\caption{\label{tab:sims} \small Simulation parameters for the simulations run down to $z=0$: simulation identifier, initial mass of gas particles ($m_\mathrm{bar}$), mass of dark matter particles ($m_\mathrm{DM}$), minimum baryonic force softening ($\epsilon_\mathrm{bar}$), minimum dark matter force softening ($\epsilon_\mathrm{DM}$), median $z=0-0.5$ stellar mass ($M_\mathrm{star}$), median $z=0-0.5$ halo mass ($M_{500c}$), median virial radius ($R_{500c}$), median SFR (M$_\odot$~yr$^{-1}$), and reference where this simulation is (or will be) described in more detail.}
\begin{tabular}[t]{llllllllll}
\hline
\hline \\[-3mm]
identifier & $m_\mathrm{bar}$ & $m_\mathrm{DM}$ & $\epsilon_\mathrm{bar}$ & $\epsilon_\mathrm{DM}$ & $M_\mathrm{star}$ & $M_{500c}$ & $R_{500c}$ & SFR & reference \\
& (M$_\odot$) & (M$_\odot$) & ($h^{-1}$pc) & ($h^{-1}$pc) & (M$_\odot$) & (M$_\odot$) & (kpc) & (M$_\odot$~yr$^{-1}$) & \\
\hline \\[-4mm]                                                                                                                                       
m10 &  $2.6\times10^2$ & $1.3\times10^3$ & 1.4 & 20 & $10^{6.3}$ & $10^{9.8}$ & 35 & $4.7\times10^{-5}$ & \citet{Hopkins2014FIRE} \\
m10v & $2.1\times10^3$ & $1.0\times10^4$ & 5 & 50 & $10^{5.6}$ & $10^{9.8}$ & 36 & $1.4\times10^{-4}$ & \citet{Ma2016} \\ 
m10h1297 & $2.1\times10^3$ & $1.0\times10^4$ & 3 & 30 & $10^{7.1}$ & $10^{10.0}$ & 43 & $1.4\times10^{-4}$ & \citet{Chan2015} \\                     
m10h1146 & $2.1\times10^3$ & $1.0\times10^4$ & 3 & 30 & $10^{8.0}$ & $10^{10.4}$ & 57 & 0.012 & \citet{Chan2015} \\ 
m10h537 & $2.1\times10^3$ & $1.0\times10^4$ & 7 & 70 & $10^{8.2}$ & $10^{10.5}$ & 63 & 0.028 & \citet{Chan2015} \\ 
m11v & $5.7\times10^4$ & $2.8\times10^5$ & 5 & 100 & $10^{9.4}$ & $10^{11.0}$ & 89 & 0.089 & \citet{Ma2016} \\ 
m11 & $7.1\times10^3$ & $3.5\times10^4$ & 5 & 50 & $10^{9.3}$ & $10^{11.1}$ & 99 & 0.048 & \citet{Hopkins2014FIRE} \\ 
m11h383 & $1.7\times10^4$ & $8.3\times10^4$ & 7 & 70 & $10^{9.5}$ & $10^{11.1}$ & 100 & 0.40 & \citet{Chan2015} \\ 
m11.4a & $3.3\times10^4$ & $1.7\times10^5$ & 6.25 & 100 & $10^{9.5}$ & $10^{11.4}$ & 122 & 0.48 & \citet{Hafen2016} \\ 
m12v & $3.9\times10^4$ & $2.0\times10^5$ & 7 & 50 & $10^{10.4}$ & $10^{11.7}$ & 159 & 2.7 & \citet{Hopkins2014FIRE} \\ 
m11.9a & $3.3\times10^4$ & $1.7\times10^5$ & 6.25 & 100 & $10^{10.1}$ & $10^{11.7}$ & 162 & 2.6 & \citet{Hafen2016} \\ 
m12i & $5.7\times10^4$ & $2.8\times10^5$ & 14 & 100 & $10^{10.6}$ & $10^{11.9}$ & 188 & 7.2 & \citet{Hopkins2014FIRE} \\ 
MFz0\_A1 & $2.7\times10^5$ & $1.4\times10^6$ & 6.25 & 100 & $10^{11.0}$ & $10^{12.7}$ & 325 & 3.7 & Feldmann et al. (in prep.) \\ 
m13 & $4.5\times10^5$ & $2.3\times10^6$ & 28 & 150 & $10^{11.0}$ & $10^{12.7}$ & 355 & 1.8 & \citet{Hopkins2014FIRE} \\ 
MFz0\_A4 & $2.7\times10^5$ & $1.4\times10^6$ & 6.25 & 100 & $10^{11.0}$ & $10^{12.7}$  & 347 & 2.0 & Feldmann et al.\ (in prep.) \\ 
MFz0\_A2 & $2.7\times10^5$ & $1.4\times10^6$ & 6.25 & 100 & $10^{11.2}$ & $10^{12.9}$ & 401 & 0.030 & \citet{Hafen2016} \\ 
\hline                                                                                                                                                
\end{tabular}
\end{center}
\end{table*}   
A $\Lambda$CDM cosmology is assumed with parameters consistent with the 9-yr Wilkinson Microwave Anisotropy Probe (WMAP) results \citep{Hinshaw2013}. The (initial) particle masses for dark matter and baryons and the minimum physical baryonic force softening length vary and are listed in Table~\ref{tab:sims} for the 16 simulations that were run down to $z=0$. The 36 high-redshift simulations that were run down to $z\approx2$ are described in \citet{Faucher2015} and \citet{Feldmann2016} and their initial baryonic (dark matter) masses range from $3.3\times10^4$ to $2.7\times10^5$~M$_\odot$ ($1.7\times10^5$ to $1.4\times10^6$~M$_\odot$). For the mass range where simulations with different resolutions overlap ($M_\mathrm{halo}=10^{10-12}$~M$_\odot$ at $z=0$ and $M_\mathrm{halo}=10^{12-13}$~M$_\odot$ at $z=2$), we find no dependence on resolution for any of the results described in this paper, which means that the hot gas properties are likely well-resolved (except perhaps for some of the X-ray emission from galactic winds, see Section~\ref{sec:concl} for a discussion). 

Star formation is restricted to molecular, self-gravitating gas above a hydrogen number density of $n_\mathrm{H}\approx10-100$~cm$^{-3}$, where the molecular fraction is calculated following \citet{Krumholz2011} and the criterion for being self-gravitating following \citet{Hopkins2013SelfGrav}. This results in the majority of stars forming at gas densities significantly higher than the imposed threshold. Stars are formed from gas satisfying these criteria at the rate $\dot\rho_\star=\rho_\mathrm{molecular}/t_\mathrm{ff}$, where $t_\mathrm{ff}$ is the free-fall time.

We assume an initial stellar mass function (IMF) from \citet{Kroupa2002}. Radiative cooling and heating are computed in the presence of the CMB radiation and the ultraviolet (UV)/X-ray background from \citet{Faucher2009}. Self-shielding is accounted for with a local Sobolev/Jeans length approximation. We impose a temperature floor of 10~K or the CMB temperature. 

The primordial abundances are $X = 0.76$ and $Y = 0.24$, where $X$ and $Y$ are the mass fractions of hydrogen and helium, respectively. 
The abundances of 11 elements (H, He, C, N, O, Ne, Mg, Si, S, Ca and Fe) produced by massive and intermediate-mass stars (through Type~Ia supernovae, Type~II supernovae, and stellar winds) are computed following \citet{Iwamoto1999}, \citet{Woosley1995}, and \citet{Izzard2004}. Mass ejected by a star particle through stellar winds and supernovae is transferred to the gas particles in its smoothing kernel.

The FIRE simulations include an explicit implementation of stellar feedback by supernovae, radiation pressure, stellar winds, and photo-ionization and photo-electric heating (see \citealt{Hopkins2014FIRE} and references therein for details). For the purposes of the present paper, we emphasize that these simulations produce galaxies with stellar masses reasonably consistent with observations over a wide range of dark matter halo masses. This is a consequence of the galactic winds driven by stellar feedback. Feedback from active galactic nuclei (AGN) is not included. Interestingly, some of the central galaxies in our simulated $10^{13}$~M$_\odot$ haloes have reasonably low SFR \citep{Voort2015b, Feldmann2016}. We focus mostly on lower-mass haloes around star-forming galaxies, where AGN are thought to be unimportant. Additionally, it is valuable to assess whether we can reproduce existing observations of hot gas around massive galaxies without AGN feedback.

We have not attempted detailed mock observations of our simulations, since we are mostly interested in scaling relations between different halo properties and are studying the mass regime where fewer detailed observations of the hot gas exist. Different ways of determining gas and halo masses can lead to biases in halo mass of 20 per cent on average \citep{Brun2014}, which is well within the scatter of the correlations we study in this paper. 

Halo mass, $M_{500c}$, is defined in this paper to be the total mass enclosed by a radius, $R_{500c}$, within which the mean overdensity is 500 times the critical density of the Universe at its redshift, $\rho_c=3H^2(z)/8\pi G$. Stellar mass, $M_\mathrm{star}$, is measured within 20~kpc of the galaxy's centre. The median halo and stellar masses from $z=0-0.5$ are given in Table~\ref{tab:sims}. Satellite galaxies are not included in this work. Throughout the paper, distances are given in proper kpc.

\subsection{Thermal Sunyaev-Zel'dovich effect} \label{sec:SZmethod}

The magnitude of the thermal SZ effect is measured by the dimensionless Compton $y$ parameter, which is proportional to the electron pressure integrated along the observer's line-of-sight:
\begin{equation} \label{eqn:y}
y = \int \sigma_T k_B T n_e / (m_e c^2) \mathrm{d}l,
\end{equation}
where $\sigma_T$ is the Thomson cross-section, $k_B$ is the Boltzmann constant, $T$ is the temperature, $n_e$ is the electron number density, $m_e$ is the electron mass, and $c$ is the speed of light.

Specifically, for each gas particle, we calculate 
\begin{equation}
\Upsilon = \sigma_T k_B T / (m_e c^2) \times m_\mathrm{gas} / (\mu_e m_H),
\end{equation}
where $m_\mathrm{gas}$ is the mass of the particle, $\mu_e$ is its mean molecular weight per free electron, and $m_H$ is the mass of a hydrogen atom. $\Upsilon$ has units of area. To calculate $Y_{500c}$, the total Comptonization parameter within radius $R_{500c}$, we sum $\Upsilon$ of all gas particles within $R_{500c}$ on the sky, $Y_{500c}$, including all particles within the simulation's zoom-in region along the line-of-sight (in order to improve comparison with observations). This is then converted to 
\begin{equation} \label{eqn:SZ}
\widetilde Y_{500c} = Y_{500c}E^{−2/3}(z)(D_A(z)/500\,\mathrm{Mpc})^2,
\end{equation}
scaled to $z=0$ and to the same angular diameter distance, $D_A$, of 500~Mpc. $E(z)$ is the Hubble parameter normalized by $H_0$, the $z=0$ Hubble parameter.

\subsection{Soft X-ray emission} \label{sec:Xmethod}

We compute $\Lambda$, the cooling function in units of erg~cm$^3$~s$^{-1}$, by interpolating a pre-computed table generated using the Astrophysical Plasma Emission Code\footnote{http://www.atomdb.org} (APEC, v1.3.1, see \citealt{Smith2001}) under the assumption that the gas is an optically thin plasma in collisional ionization equilibrium. APEC cooling rates are computed for each element individually as a function of photon energy on an element-by-element basis. The total cooling rate is calculated by summing over the 11 most important elements for cooling (hydrogen, helium, carbon, nitrogen, oxygen, neon, magnesium, silicon, sulphur, calcium and iron) which are tracked during the simulation. APEC assumes the solar abundance ratios of \citet{Anders1989}, but we modified the spectra to reflect the actual simulated abundances.

$\Lambda$ is derived from the tables as a function of $\mathrm{log}_{10}T$ through log-linear interpolation and is integrated over the energy band of interest, in this paper $0.5-2.0$~keV. The contribution of element $j$ to a gas particle's luminosity is then
\begin{equation} \label{eqn:X}
L_X =  \Lambda(T) n_e n_\mathrm{H} \dfrac{m_\mathrm{gas}}{\rho} \dfrac{X_j}{X_j^\odot},
\end{equation}
where $\rho$ is the particle's density, $X_j$ is the mass fraction of the element and $X_j^\odot$ is the solar mass fraction of the same element. The soft X-ray emission from gas below $10^6$~K is negligible. 

The simulations do not contain stellar point sources, such as X-ray binaries, and therefore all X-ray emission is diffuse emission. It should be noted that the APEC cooling tables, which assume pure collisional ionization equilibrium, neglect the extragalactic UV/X-ray background. However, the effect of photo-heating on the derived X-ray properties is small in the regime we explore here \citep{Wiersma2009a, Crain2010a} and the APEC cooling rates are therefore consistent with those used for running the simulations.

\section{Results} \label{sec:results}

\begin{figure}
\center
\includegraphics[scale=.55]{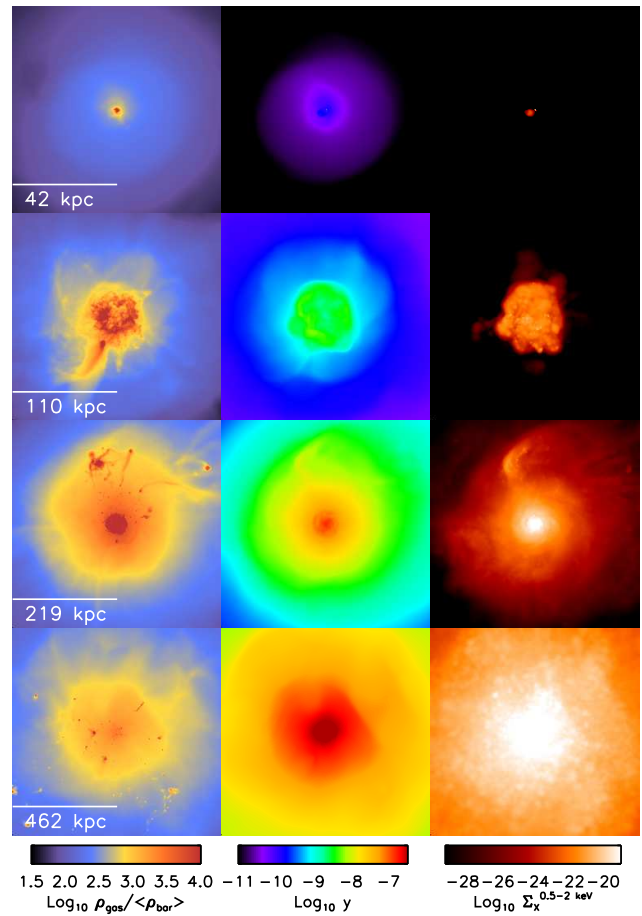}
\caption {\label{fig:img} Images of the gas density, Compton $y$ parameter (SZ signal; see Eqn.\ \ref{eqn:y}), and soft X-ray surface brightness (in erg~s$^{-1}$~cm$^{-2}$~arcsec$^{-2}$) at $z=0.125$ ($D_A\approx500$~Mpc) out to $R_{500c}$ for 4 representative haloes. The value of $R_{500c}$ is indicated in the first column. From top to bottom the halo mass increases by about an order of magnitude in each row, from $10^{9.8}$ to $10^{12.9}$~M$_\odot$. The second row shows that stellar feedback can result in strong X-ray emission.}
\end{figure}
Figure~\ref{fig:img} shows the gas density scaled by the average density of the Universe, $\langle\rho_\mathrm{bar}\rangle$, the SZ effect as measured by the (dimensionless) $y$ (Equation~\ref{eqn:y}), and the soft X-ray surface brightness, $\Sigma_X^\mathrm{0.5-2\,keV}$ (in erg~s$^{-1}$~cm$^{-2}$~arcsec$^{-2}$) for 4 haloes of different masses (m10, m11, m12i, MFz0\_A2) at $z=0.125$ ($D_A\approx500$~Mpc). 
The halo mass increases by about an order of magnitude in each row, from top ($M_\mathrm{halo}=10^{9.8}$~M$_\odot$) to bottom ($M_\mathrm{halo}=10^{12.9}$~M$_\odot$). The size of each image is $2R_{500c}$ by $2R_{500c}$ and therefore increases from top (84~kpc) to bottom (924~kpc) as well. The disturbed morphology and sharp edge to the X-ray flux visible in the second row hints at the fact that stellar feedback plays an important role and that outflows can power X-ray emission in low-mass objects.

\begin{figure}
\center
\includegraphics[scale=.52]{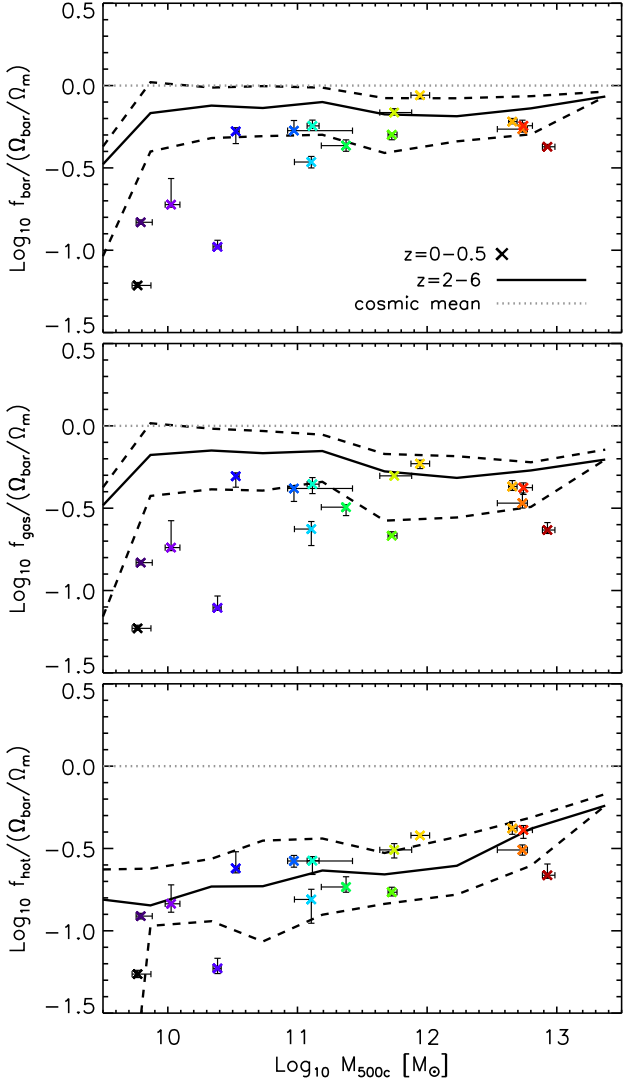}
\caption {\label{fig:frac} Baryon fraction (top), gas fraction (middle), hot gas fraction ($T>10^4$~K; bottom) normalized by the cosmic baryon fraction within $R_{500c}$ as a function of halo mass. The black curves show median (solid) and 16th and 84th percentiles (dashed) of the fractions at $z\ge2$. The crosses show the median values for $z=0-0.5$, with colours varying with halo mass. The error bars indicate the $1\sigma$ scatter in both the fractions and halo mass over this redshift interval. The baryon and (hot) gas fractions are below the cosmic mean, especially at low redshift and for low $M_{500c}$. The fraction of hot (ionized) gas is primarily a function of halo mass, with little dependence on redshift.}
\end{figure}
%.
Feedback changes the properties of both galaxies and their haloes by expelling gas from the interstellar medium (ISM). This in turn enriches and heats the circumgalactic medium or halo gas. These galactic winds can be strong enough to remove gas from the haloes of galaxies, especially for low-mass haloes \citep[e.g.][]{Muratov2015}. 
In Figure~\ref{fig:frac} we show, from top to bottom, the baryon fraction, gas fraction, and hot gas (above $10^4$~K) fraction normalized by the cosmic baryon fraction ($\Omega_\mathrm{bar}/\Omega_\mathrm{m}=0.16$) for the full suite of FIRE simulations as a function of $M_{500c}$. The black curves show the median (solid) and 16th and 84th percentiles (dashed) of these fractions at $z=2,3,4,5$, and 6 combined (this is done to increase statistics and enable comparison with Figure~\ref{fig:Y_mass} later on). The coloured crosses show median values in the redshift range $z=0-0.5$, with associated error bars indicating the 16th and 84th percentiles (i.e.\ the scatter associated with time variability). The scatter for individual objects is relatively small, since galaxies grow only slowly at low redshift, except when they experience a major merger. 

The baryon fraction and gas fraction are reduced compared to the cosmic mean baryon fraction. This effect is larger at low halo mass, where it is easier for stellar feedback to expel gas from the halo \citep[e.g.][]{Chan2015, Badry2016}. The gas and baryon fractions are also lower at lower redshift, after powerful outflows at intermediate redshift $z\approx0.5-2$ remove a large amount of gas from the halo \citep{Muratov2015}. The hot, ionized gas fraction is necessarily the lowest. Interestingly, it is also the only one where the high-redshift and low-redshift results are consistent with each other. A much larger fraction of the gas is cold ($<10^4$~K) at high redshift and a smaller fraction of the baryons have been expelled, but these effects cancel out and lead to similar $f_\mathrm{hot}$. At low redshift, haloes with $M_{500c}\approx10^{11-12}$~M$_\odot$ show the biggest difference between total gas and hot gas fraction, which is also the mass regime where one expects to find large cold gas reservoirs and high SFRs. 

The (hot) gas fractions of groups and clusters have been derived from X-ray observations by many different groups and they are found to increase with halo mass \citep[e.g.][]{Vikhlinin2009, Sun2009, Gonzalez2013}. 
Our simulated galaxy groups have gas fractions consistent with those observed. There is large scatter between haloes in observations, which we are not able to fully probe with our limited simulation sample, though there is also clear evidence for scatter between objects at fixed halo mass in Figure~\ref{fig:frac} (e.g.\ around $10^{13}$~M$_\odot$).

\subsection{Thermal Sunyaev-Zel'dovich signal} \label{sec:SZ}

\begin{figure}
\center
\includegraphics[scale=.52]{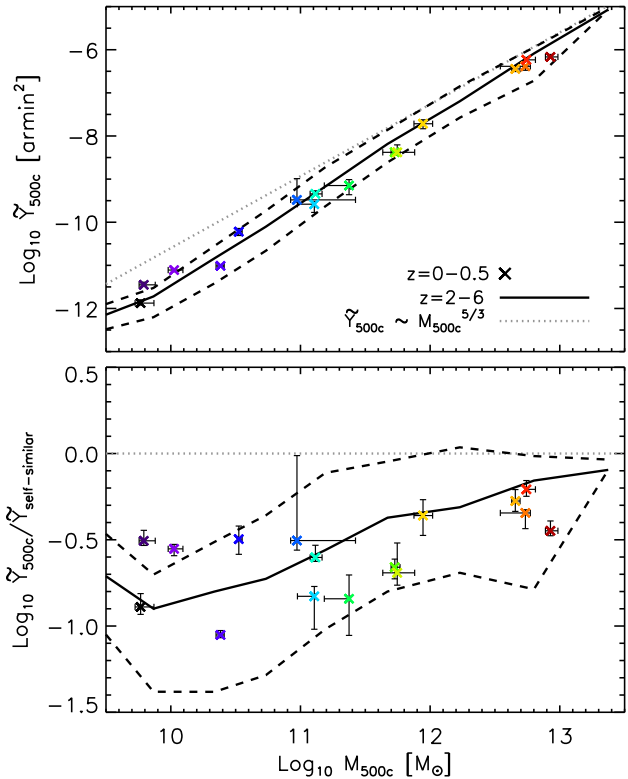}
\caption {\label{fig:Y_mass} The Sunyaev-Zel'dovich signal as measured by $\widetilde{Y}_{500c}$ within a 2D radius $R_{500c}$ as a function of $M_{500c}$. The black curves show median (solid) and 16th and 84th percentiles (dashed) of $\widetilde{Y}_{500c}$ at $z\ge2$. The coloured crosses show the median values for $z=0-0.5$, with the error bars indicating the 16th and 84th percentiles of $\widetilde{Y}$ and $M_{500c}$ over the same redshift interval. The dotted grey line shows the self-similar relation $\widetilde{Y}_{500c}\propto M_{500c}^{5/3}$ (taken from \citet{Planck2013}, extrapolated below $M_{500c}=10^{12.6}$~M$_\odot$). In the bottom panel the measured $\widetilde{Y}$ is divided by the Planck fit. There is a clear reduction in the SZ signal, compared to self-similar, which becomes stronger towards lower halo masses in a way similar to the behaviour of the hot gas fraction in Figure~\ref{fig:frac}.}
\end{figure}
Figure~\ref{fig:Y_mass} shows $\widetilde{Y}_{500c}$ (defined in Equation~\ref{eqn:SZ}) within a 2D radius $R_{500c}$ as a function of halo mass for both high (black curves) and low (coloured crosses with error bars) redshift. The symbols and linestyles show the same as in Figure~\ref{fig:frac}, i.e.\ median values and 16th and 84th percentiles. $\widetilde{Y}$ is independent of redshift as it scales out the dependence on cosmology and puts all objects at a distance $D_A=500$~Mpc. We include all the gas along the line-of-sight in our high-resolution simulation region, but in practice this is very similar to integrating out to a few times $R_{500c}$. The dotted, grey line in the top panel is a self-similar fit (with a slope\footnote{The mass contributing to $\widetilde{Y}_{500c}$ increases linearly with $M_{500c}$ and the virial temperature scales as $T_\mathrm{vir}\propto M_{500c}^{2/3}$, therefore $\widetilde{Y}_{500c}\propto M_{500c}^{5/3}$.} of 5/3) based on stacks of observed locally brightest galaxies \citep{Planck2013}, which we extrapolated below $M_{500c}\approx10^{12.6}$~M$_\odot$. We find a larger suppression of the SZ signal towards lower masses. The best fit to our simulation results over the full mass range gives $\widetilde{Y}_{500c}\propto M_{500c}^{1.8}$, somewhat steeper than self-similar. 

This is shown more clearly in the bottom panel of Figure~\ref{fig:Y_mass}, where we divide the SZ signal from our simulations by the self-similar relation. The signal is reduced by up to an order of magnitude for our smallest dwarf haloes. This is directly related to the stronger decrease in the hot, ionized gas fraction at lower masses shown in Figure~\ref{fig:frac}, since neutral gas does not contribute to $n_e$ and thus to $\widetilde{Y}$. Note, however, that the $f_\mathrm{hot}$ measurement is made in 3D. Our SZ signal when measured within a 3D radius is slightly lower (by $0.1-0.2$~dex) than in 2D, because hot, ionized gas outside $R_{500c}$ contributes to the measurement as well, especially for low-mass haloes. 

\begin{figure}
\center
\includegraphics[scale=.52]{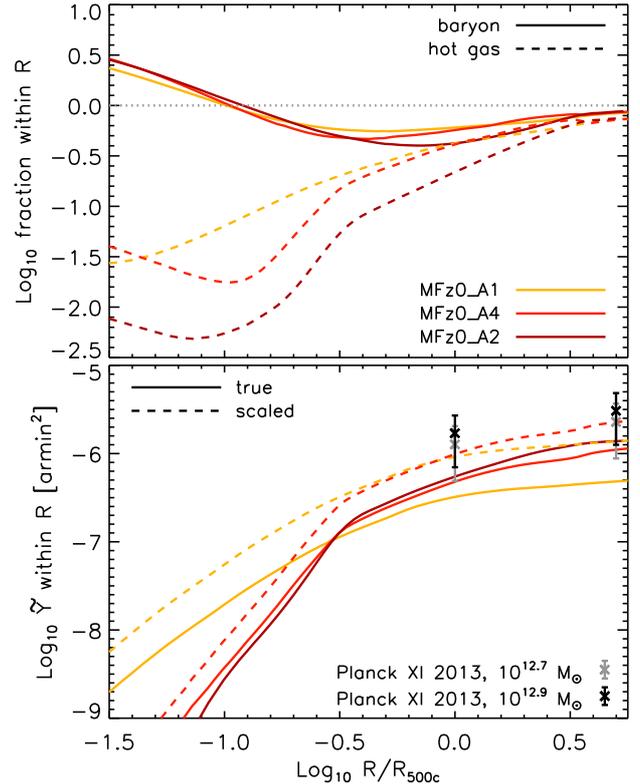}
\caption {\label{fig:R} 3D radial profiles of three massive haloes (MFz0\_A*) at $z=0.25$ as a function of $R/R_{500c}$. Top panel: Baryon fraction (solid) and hot gas fraction ($T>10^4$~K; dashed) normalized by the cosmic baryon fraction within radius $R$. The horizontal dotted line shows the cosmic baryon fraction. Outside the virial radius the baryon and gas fractions increase and approach the cosmic mean at $5R_{500c}$. Bottom panel: $\widetilde{Y}$ within $R$ (solid curves). Dashed curves show $\widetilde{Y}$ for the two lower-mass haloes ($10^{12.7}$~M$_\odot$) scaled to the mass of the most massive halo ($10^{12.9}$~M$_\odot$) by assuming self-similarity. Crosses with 1$\sigma$ error bars show the measured (at $5R_{500c}$) and inferred (at $R_{500c}$) average $\widetilde{Y}$ \citep{Planck2013}. Although $\widetilde{Y}$ increases with radius, the discrepancy with the observed SZ signal does not decrease significantly.}
\end{figure}

For galaxies with $M_{500c}>10^{12.6}$~M$_\odot$ (or $M_\mathrm{star}>10^{10.9}$~M$_\odot$), where our SZ predictions overlap with the stacked \citet{Planck2013} observations, we find a modest factor of $\sim2$ deficit of the SZ signal relative to their results. This can potentially be explained by the large angular resolution of the observations. The SZ signal with Planck is measured at a much larger radius ($5R_{500c}$), within which the baryon fraction is closer to the cosmic mean, and then rescaled (by a factor of 1.796). \citet{Brun2015} show that following the same approach brings their simulations into agreement with the observations even though the SZ signal within $R_{500c}$ appears lower than observed. In Figure~\ref{fig:R} we investigate whether that is true for our massive galaxies as well. 

The top panel of Figure~\ref{fig:R} shows hot gas and baryon fractions (as in Figure~\ref{fig:frac}) as a function of radius, out to $5R_{500c}$, for three of our massive haloes (MFz0\_A1, MFz0\_A4, MFz0\_A2). The colours are consistent with those used in the other figures. The fourth massive halo (m13) is not shown, because its simulated volume only goes out to $R\approx R_{500c}$. Solid curves show $f_\mathrm{bar}$ and dashed curves $f_\mathrm{hot}$. 
The cosmic baryon fraction is indicated by the horizontal dotted line. 
The baryon fraction inside $0.1R_{500c}$ is higher than the cosmic mean, because baryons are able to cool and therefore reach smaller radii than the dark matter in the halo. The baryon fraction decreases with radius until it reaches a broad minimum, below the cosmic mean, around the virial radius. This minimum is set by galactic winds pushing gas out to beyond virial radius. At $R\gtrsim R_{500c}$ the baryon fraction increases and approaches the cosmic mean at $5R_{500c}$. The hot gas fraction is much lower than cosmic in the centre of the halo, but approaches the baryon fraction outside the virial radius. The total gas fraction is very close to the hot gas fraction at this halo mass. The baryon and hot gas fractions are within 0.15~dex of the cosmic baryon fraction at $5R_{500c}$. This behaviour is qualitatively similar for the other lower-mass haloes. 

\begin{figure*}
\center
\includegraphics[scale=.52]{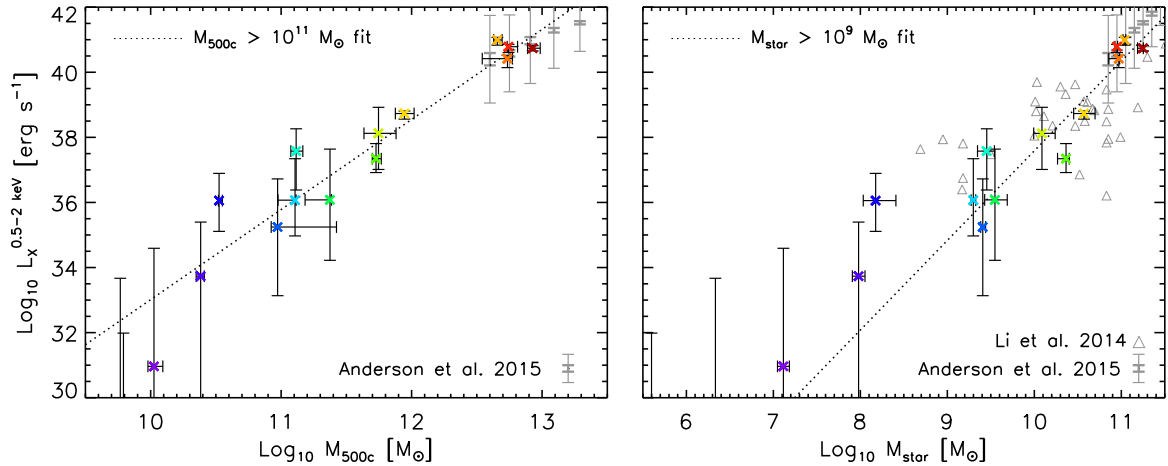}
\caption {\label{fig:X_mass} The median soft X-ray luminosity ($0.5-2$~keV) within 2D radius $R_{500c}$ from $z=0.5-0$ (coloured crosses) as a function of halo mass (left panel) and stellar mass (right panel) for each halo. The associated error bars show the 16th and 84th percentiles ($1\sigma$ scatter) over the same redshift interval. Stacked observations from \citet{Anderson2015} are shown by grey error bars, with the thick error bars showing the error on the measurement and the thin error bars showing the upper limit of the intrinsic scatter. Grey triangles in the right panel show measurements from \citet{Li2014}, which only probe out to $0.1R_{200c}$. The black, dotted lines show the best fit for $M_{500c}>10^{11}$~M$_\odot$ (left) and $M_{star}>10^{9}$~M$_\odot$ (right). Our simulations are in agreement with \citet{Anderson2015}, but somewhat discrepant from \citet{Li2014} (see text). In our simulations, the scatter in $L_X$ increases towards lower masses, indicating that the emission becomes more time variable.}
\end{figure*}

The bottom panel of Figure~\ref{fig:R} shows $\widetilde{Y}$ as a function of radius as solid curves. Dashed curves show the same profile rescaled to the mass of the most massive halo ($10^{12.9}$~M$_\odot$) assuming self-similarity ($\widetilde{Y}\propto M^{5/3}$). Variations in the temperature profiles of the haloes, as well as in $f_\mathrm{hot}$, are responsible for the differences in $\widetilde{Y}$ at small radii between different haloes. The black (grey) crosses and 1$\sigma$ error bars show the values derived from \citet{Planck2013} observations for $10^{12.9}$~M$_\odot$ ($10^{12.7}$~M$_\odot$) at $R_{500c}$ and $5R_{500c}$ (multiplied by 1.796). The two simulated lower-mass haloes should be compared to the grey data points (when not scaled). At both $R_{500c}$ and $5R_{500c}$ our haloes fall a factor of a few below the observed mean $\widetilde{Y}$, but are within $2\sigma$. 
It is also important to stress that a reanalysis of the same Planck data by \citet{Greco2015} finds larger error bars at $M_\mathrm{star}\approx10^{11.3}$~M$_\odot$ and only upper limits at $M_\mathrm{star}\lesssim10^{11.1}$~M$_\odot$ ($M_\mathrm{500c}\lesssim10^{13}$~M$_\odot$), so our simulations are fully consistent with their measurements. Stacking haloes in a mass bin and taking the average flux will necessarily bias the measurement towards high values, so it is possible that in a large statistical sample, the mean $\widetilde{Y}$ in our simulations would be somewhat higher. %reproduce the mean value of the \citet{Planck2013} observations.

\subsection{Soft X-ray luminosity} \label{sec:Xray}

Because diffuse X-ray emission around galaxies is generally measured at low redshift, we only show results from simulations that were run down to $z=0$ in this section. Although we show results averaged from $z=0.5-0$ to reduce stochastic effects associated with time variability in low-mass systems, they can be compared directly to $z=0$ observations, because there is no trend in our simulations with redshift at $z<0.5$. This is due to the fact that the mass of the systems does not increase significantly and because there is no systematic evolution in the star formation rate. 

% figure

The relation between soft X-ray luminosity and halo mass is shown in Figure~\ref{fig:X_mass}. The coloured crosses show the median soft X-ray luminosity between $z=0$ and $z=0.5$ for each of our simulations as a function of median halo mass (left panel) and median stellar mass (right panel). Our simulations show a very steep scaling of $L_X\propto M_{500c}^{2.7}$ (for $10^{11}<M_{500c}<10^{13}$~M$_\odot$) and $L_X\propto M_\mathrm{star}^{2.7}$ (for $10^9<M_\mathrm{star}<10^{11.5}$~M$_\odot$) as shown by the black, dotted lines. This is much steeper than a self-similar relation with slope 4/3, which assumes that the X-ray luminosity is dominated by thermal bremsstrahlung \citep{Sarazin1986} and also steeper than the slope found in observations, 1.84 \citep{Anderson2015}. This observational work, however, probes larger halo masses than we do here. 

\begin{figure}
\center
\includegraphics[scale=.52]{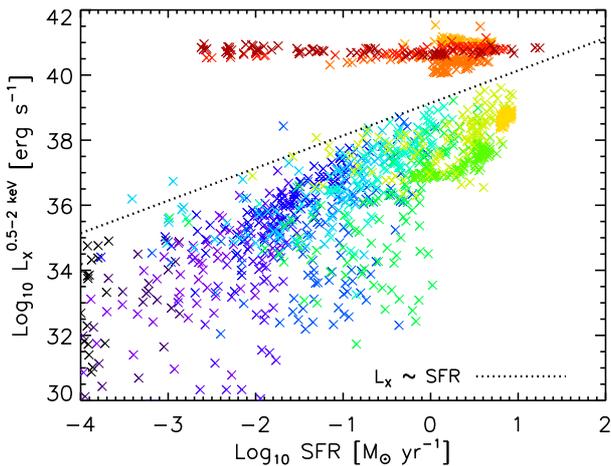}
\caption {\label{fig:X_SFR} The soft X-ray luminosity ($0.5-2$~keV) as a function of SFR (averaged over 100~Myr) for each individual galaxy halo from $z=0-0.5$ (at 100 approximately equally spaced times). Crosses with the same colour belong to the same galaxy at different times. Colours correspond to different halo masses and match the colours used in previous figures. The dotted line shows the linear $L_X-$SFR relation for diffuse gas found in observations by \citet{Li2013b}, which we extrapolated below SFR$=0.01$~M$_\odot$~yr$^{-1}$. For halo masses $\lesssim10^{12}$~M$_\odot$, $L_X$ correlates strongly with SFR, whereas for $M_{500c}\gtrsim10^{13}$~M$_\odot$ $L_X$ is independent of SFR (see Figure~\ref{fig:X_slope}). This shows the different origin of the diffuse X-ray emission: for lower-mass haloes it is powered by galactic winds, whereas for higher-mass haloes it originates from gas heated by an accretion shock at the virial radius.}
\end{figure}

\begin{figure}
\center
\includegraphics[scale=.52]{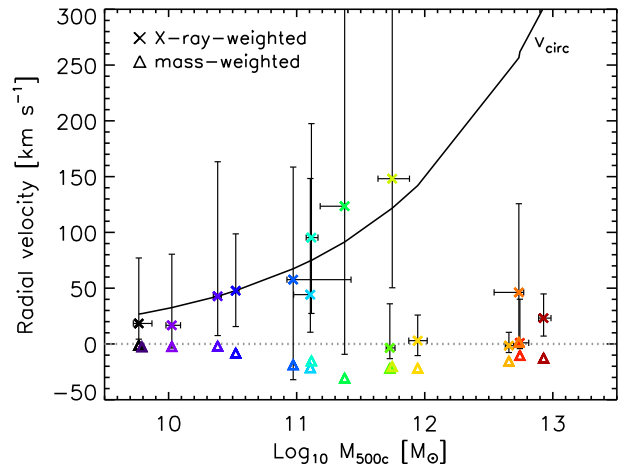}
\caption {\label{fig:X_vrad} Median mass-weighted (triangles) and soft X-ray-weighted (crosses) radial velocity of hot, diffuse gas ($T>10^4$~K) as a function of halo mass at $z=0-0.5$. The error bars show the 16th and 84th percentiles for the X-ray-weighted velocity. The solid, black curve shows the circular velocity, $v_\mathrm{circ}$, of the haloes at $R_{500c}$. The grey, dotted curve indicates a static halo. Soft X-ray emission is highly biased towards outflows for low-mass systems ($M_{500c}\lesssim10^{11.5}$~M$_\odot$), for which the X-ray weighted radial velocity is close to $v_\mathrm{circ}$, on average. For high-mass objects ($M_{500c}\gtrsim10^{12.5}$~M$_\odot$) the X-ray-weighted radial velocity is low compared to $v_\mathrm{circ}$, indicating that the emission comes from the virialized halo gas.}
\end{figure}

The error bars on the simulation data points in Figure~\ref{fig:X_mass} indicate the 16th and 84th percentile (or scatter) over the same redshift interval and therefore show the variation of the X-ray luminosity over time. This time variation is very small for galaxy groups, but increases dramatically and steadily towards lower masses. This is because at lower halo masses, hot galactic outflows driven by massive stars contribute to the total X-ray luminosity, increasing the stochasticity. This is further explored below, in Figures~\ref{fig:X_SFR} and~\ref{fig:X_vrad}. 

The grey error bars in Figure~\ref{fig:X_mass} are measurements from \citet{Anderson2015} who stacked locally brightest galaxies and included all soft X-ray emission within R500c. The thick error bars show the error on the mean derived from bootstrapping, whereas the thin error bars show an upper limit on the amount of intrinsic scatter\footnote{This is an upper limit, because it contains the intrinsic scatter plus unknown contributions from X-ray binaries and low-luminosity AGN.} between different haloes. The X-ray luminosity in our simulations is measured in 2D for consistency with observations, but we checked that this is very similar to 3D since the emission scales as $\rho^2$ and is thus highly centrally concentrated. Our simulations are consistent with the stacked observations in the mass range where they overlap. 

In the right panel, the grey triangles show observations of field galaxies from \citet{Li2014}, which probe out to smaller radii ($0.01<R/R_\mathrm{200c}<0.1$) than our simulations and \citet{Anderson2015}. We have checked that this smaller radius changes $L_X$ by about $\sim0.5$~dex on average. At low stellar masses, our simulations predict X-ray luminosities that are of order, although a bit lower than, the observations. A reason for this could be that the observational data still contain a contribution from unresolved X-ray point sources or that the observational sample is biased towards high $L_X$, because of the way the objects are selected. Alternatively, our simulations could be missing X-ray emission arising from the interfaces between hot halo gas and cool clouds, as discussed in \ref{sec:concl}. At high stellar masses, our simulations predict X-ray luminosities that are above most of the \citet{Li2014} measurements. This discrepancy is partially due to the smaller radii probed in the observations, which therefore miss a substantial fraction of the total X-ray emission. Additionally, it is possible that star-forming galaxies live in somewhat lower-mass haloes than quiescent galaxies with the same stellar mass \citep{Mandelbaum2016}, in which case they would have lower virial temperatures and thus lower X-ray luminosities. 

The scatter in $L_X$ for low-mass haloes in our simulations is large, but many galaxies with $M_\mathrm{star}\gtrsim10^9$~M$_\odot$ have $L_X>10^{37}$~erg s$^{-1}$ at least part of the time, which means they would be detectable with current instruments \citep[e.g.][]{Li2013b}. In an unbiased survey, our simulations predict a large number of sources with lower $L_X$, which may be currently undetectable. 

\begin{figure}
\center
\includegraphics[scale=.52]{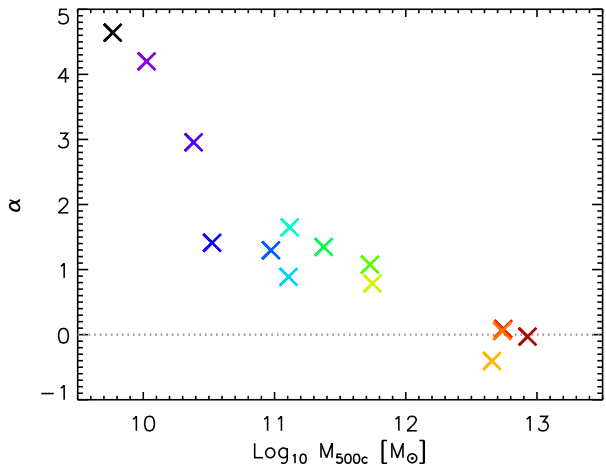}
\caption {\label{fig:X_slope} The power, $\alpha$, of the correlation between $L_X$ and SFR ($L_X\propto$~SFR$^\alpha$) for each individual halo as a function of halo mass. The dotted line at $\alpha=0$ corresponds to no correlation. m10v and m12i have been omitted from this figure, because $\alpha$ could not reliably be determined. The X-ray emission around dwarf galaxies is a strong function of their SFR, while haloes with $M_{500c}\approx10^{11-12}$~M$_\odot$ exhibit a correlation close to linear. There is, however, no correlation between $L_X$ and SFR for the most massive objects, because hot, virialized halo gas produces more X-rays than star formation powered winds.}
\end{figure}

Both the SFR (averaged over 100~Myr) and the soft X-ray luminosity ($0.5-2$~keV) vary strongly with time. Figure~\ref{fig:X_SFR} shows $L_X$ as a function of SFR for each of our simulations for 100 snapshots between $z=0.5$ and 0. Crosses with the same colour belong to the same galaxy at different times. Colours correspond to different halo masses and match the colours used in previous figures. 
For halo masses $\lesssim10^{12}$~M$_\odot$ $L_X$ correlates strongly with SFR, whereas for $M_{500c}\gtrsim10^{13}$~M$_\odot$ $L_X$ is completely independent of SFR (see Figure~\ref{fig:X_slope}). This difference is caused by the different origin of the diffuse X-ray emission, which is powered by galactic winds for lower-mass haloes and by an accretion shock at the virial radius for higher-mass haloes. Note that the SFRs of our most massive objects lie at the high end of what is observed for local early-type galaxies \citep{Davis2014}. However, decreasing their SFRs would not affect $L_X$ and only strengthen our conclusion that the X-ray emission originates from shock-heated, virialized halo gas.
There is significant scatter at fixed SFR for low halo masses. This is at least in part due to the fact that there will be some time delay between the formation of new stars and supernova-driven outflows. A similar correlation between SFR and luminosity has also been found for ultra-violet metal-line emission at high redshift \citep{Sravan2016}. 

The dotted line in Figure~\ref{fig:X_SFR} shows the observed linear correlation between the total diffuse X-ray luminosity and SFR for a (heterogeneous) sample of star-forming galaxies with $\mathrm{SFR}>0.01$~M$_\odot$~yr$^{-1}$ \citep{Li2013b}\footnote{This is similar to the linear SFR-X-ray correlation in \citet{Mineo2012}, though a factor of $\sim1.7$ higher in normalization, because the observations of \citet{Li2013b} probe out to larger radii.}. The relation is extrapolated below a SFR of 0.01~M$_\odot$~yr$^{-1}$. 
In general there is a reasonably good match between our simulations and observations. However, at fixed SFR, the average $L_X$ in our simulations is somewhat lower than that observed, especially at low SFR. This could be explained if the observations still include a contribution from X-ray binaries or are biased towards high $L_X$, because of the way the sample is selected. Alternatively, our simulations may be underpredicting $L_X$ from galactic winds (see Section~\ref{sec:concl}). 

The $L_X-$SFR correlation for our ensemble of simulated galaxies is somewhat steeper than linear. However, if we restrict ourselves to $L_X>10^{36}$~erg~s$^{-1}$ and SFR$\,>0.01$~M$_\odot$~yr$^{-1}$, imitating possible observational detection limits, the slope of the correlation is $\sim0.8$, close to linear. 

Figure \ref{fig:X_SFR} shows that the X-ray luminosity of the three objects with $M_{500c}\approx10^{12}$~M$_\odot$ is never below $10^{36}$erg~s$^{-1}$ even at low SFRs of less than 0.1~M$_\odot$~yr$^{-1}$. This is likely because the gas that went through an accretion shock at the virial radius sets a lower limit for $L_X$. For these haloes ($M_{500c}\approx10^{12}$~M$_\odot$) it is possible that in high-SFR systems the X-rays are powered (directly or indirectly) by galactic winds, whereas in low-SFR systems the X-rays originate from the diffuse, quasi-hydrostatic halo gas. 
Our simulated sample does not contain any strongly starbursting massive galaxies at low redshift. If the scaling with SFR seen in lower-mass star-forming galaxies persists, we would expect galaxies with $M_\mathrm{star}=10^{11}$~M$_\odot$ or $M_\mathrm{halo}=10^{13}$~M$_\odot$ to become dominated by X-rays powered by galactic outflows if their SFR$\gtrsim100$~M$_\odot$~yr$^{-1}$ (e.g.\ (ultra)luminous infrared galaxies). 

Another clear distinction between the X-ray emission of our most massive haloes and those around star-forming galaxies lies in the radial profile. The surface brightness of the latter falls off more steeply with radius than that of the former, as can also be seen in Figure~\ref{fig:X_prof}.

A correlation between $L_X$ and SFR for an ensemble of star-forming galaxies is expected because both properties increase with galaxy (and halo) mass. 
\citet{Wang2016} show that, indeed, most of the $L_X-$SFR correlation is due to the differences in stellar mass and when they scale this out, the correlation becomes very sub-linear. Theoretically, one also expects a correlation between $L_X$ and SFR for objects of the \emph{same} mass if supernovae drive hot winds out of star-forming galaxies. Such a correlation at fixed stellar mass would therefore be the most direct observational test of the importance of stellar feedback. 

To scale out the dependence on stellar mass, Figure~\ref{fig:X_slope} shows the slope, $\alpha$, of the correlation between $L_X$ and SFR  ($L_X\propto$~SFR$^\alpha$) from $z=0-0.5$ as a function of median halo mass for each individual halo. We have excluded simulations m10v and m12i for which the correlation could not reliably be determined, because m10v has too few nonzero $L_X$ data points and m12i has too little variation in SFR. 

Galaxies in the smallest haloes ($M_{500c}\approx10^{10}$~M$_\odot$) show correlations much steeper than linear, up to a power of 5. Galaxies in haloes of $M_{500c}\approx10^{11-12}$~M$_\odot$ show correlations close to linear. For galaxies in the most massive haloes ($M_{500c}\approx10^{13}$~M$_\odot$), there is no correlation between SFR and X-ray luminosity. This is due to the fact that the hot, hydrostatic halo gas (with temperatures close to the virial temperature) dominates the X-ray emission and the energy input from stellar feedback is insignificant compared to the energy input by the accretion shock at the virial radius. The strong dependence of $\alpha$ on halo mass indicates that the importance of galactic winds for the X-ray emission decreases smoothly with increasing halo mass. 

To better assess the dynamical state of the X-ray emitting gas, we calculate the average mass-weighted and soft X-ray-weighted radial velocity of hot, diffuse gas with $T>10^4$~K and $n_\mathrm{H}<0.1$~cm$^{-3}$ (thus excluding the ISM). Figure~\ref{fig:X_vrad} shows the median mass-weighted (triangles) and X-ray-weighted (crosses) radial velocity as a function of halo mass, as well as the 16th and 84th percentiles for the X-ray-weighted velocity from $z=0-0.5$. There is no inherent reason why the emission would be biased towards gas with high radial velocity, yet the average X-ray-weighted radial velocity is close to the halo's circular velocity, $v_\mathrm{circ}$, for $M_{500c}\lesssim10^{11.5}$~M$_\odot$, albeit with large scatter. This means that the X-ray emission is much more likely to be associated with outflowing gas than the average hot halo gas. For high-mass haloes, $M_{500c}\gtrsim10^{12.5}$~M$_\odot$, the X-ray-weighted radial velocity is much lower than the halo's $v_\mathrm{circ}$ and only slightly higher than the mass-weighted radial velocity. In these haloes the X-ray emission is dominated by quasi-hydrostatic halo gas.

Some of our Milky Way-mass haloes ($M_{500c}\approx10^{12}$~M$_\odot$) show relatively low X-ray-weighted radial velocities, whereas another shows velocities as high as $v_\mathrm{circ}$. This is an indication that for some haloes, outflows are still contributing to their X-ray emission, but that for others the halo gas heated by an accretion shock is becoming more important. Milky Way-mass galaxies are close to, but still above, the transition mass below which haloes no longer exhibit a stable virial shock \citep[e.g.][]{Birnboim2003}. In these haloes the two sources for diffuse X-rays compete: when the SFR and outflow rate are low, the halo gas heated through a virial shock dominates, whereas when the SFR is high, the hot galactic wind and halo gas shocked by the wind dominate. This is also consistent with the $L_X-$SFR correlation becoming shallower with increasing mass (see Figure~\ref{fig:X_slope}).

X-ray emission is naturally biased towards high density, temperature, and metallicity \citep[e.g.][]{Voort2013, Crain2013}. We find the same bias in our simulations. The soft X-ray-weighted metallicity and density are about an order of magnitude higher than the mass-weighted ones (for gas with $T>10^4$~K). The X-ray-weighted metallicities of the $M_{500c}\approx10^{13}$~M$_\odot$ objects are close to solar, but their mass-weighted metallicities are only $\sim0.1$~Z$_\odot$. For lower-mass galaxies, the X-ray-weighted (mass-weighted) metallicities are about $0.3-0.5$~Z$_\odot$ ($0.03-0.1$~Z$_\odot$). There is no significant change in this bias across the full mass range we probe. This is not the case for the temperature bias, which decreases from 1.4~dex for dwarf galaxies to 0.3~dex for our group-sized haloes. We therefore conclude that the outflow bias seen in Figure~\ref{fig:X_vrad} is driven by temperature differences. In low-mass haloes all hot, X-ray emitting halo gas is contained in outflows driven by stellar feedback, whereas at higher halo masses the hot gas is dominated by accreted gas that was shock-heated at the virial radius. Stellar feedback becomes less important for the X-ray luminosity towards higher masses, because the virial temperature increases and thus the contribution from hydrostatic halo gas also increases.

Note that our simulations include only stellar feedback. It is possible that AGN feedback would heat the halo gas and increase the X-ray-weighted radial velocity in the most massive objects. However, it would have to do so without significantly changing $L_X$ in order to stay in agreement with observations.

\section{Discussion and conclusions} \label{sec:concl}

In this paper, we have quantified the baryon and gas fractions, the strength of the Sunyaev-Zel'dovich (SZ) effect, and the soft X-ray luminosity of galaxy haloes using a large suite of zoom-in simulations from the FIRE project. These observational probes reveal the impact of strong stellar feedback on gaseous haloes (our simulations do not include AGN feedback). Consistent with previous results, we find that lower-mass galaxies are more affected by galactic winds driven by stellar feedback. 

We find a relation between the integrated electron pressure, as measured by the SZ effect, and halo mass, $M_{500c}$: $\widetilde{Y}_{500c}\propto M_{500c}^{1.8}$ for $M_{500c}\lesssim10^{13}$~M$_\odot$ (Fig.\ \ref{fig:Y_mass}; see Equation~\ref{eqn:SZ} for the definition of $\widetilde{Y}_{500c}$). 
This is steeper than the self-similar scaling of $\widetilde{Y}_{500c}\propto M_{500c}^{5/3}$, implying that the SZ effect is more strongly suppressed by stellar feedback within the virial radius of lower-mass haloes. This is driven by the decreasing fraction of ionized gas with decreasing halo mass (Fig.\ \ref{fig:frac}). At low redshift, the fraction of ionized gas is strongly correlated with the total gas fraction and baryon fraction, which also decrease with decreasing halo mass, owing to more efficient expulsion of gas in lower-mass haloes. At high redshift, galaxies are rich in cold, star-forming gas and have baryon fractions much closer to universal, yet this gas does not contribute to the SZ effect, because it is neutral. The resulting value of $\widetilde{Y}_{500c}$ as a function of mass is remarkably independent of redshift.

Our simulated SZ values (at $M_{500c}\approx10^{13}$~M$_\odot$) are a few times lower than those measured by \citet{Planck2013}, both at low and high redshift, but still within $2\sigma$. However, \citet{Greco2015} find a larger noise contribution using the same data and only upper limits at this halo mass, which are fully consistent with our simulations. The \citet{Planck2013} observations probe a scale about five times large than $R_{500c}$, within which the baryon fraction is close to the cosmic mean (Fig.\ \ref{fig:R}). The resulting $\widetilde{Y}$ is then rescaled by assuming a `universal pressure profile', based on measurements of galaxy clusters. \citet{Brun2015} use full cosmological simulations (with lower resolution, better galaxy statistics, and AGN feedback) to predict a large reduction of $\widetilde{Y}_{500c}$ for $M_{500c}\approx10^{13}$~M$_\odot$ at $R_{500c}$ (about a factor~5) and a value consistent with \citet{Planck2013} at $5R_{500c}$. This is fundamentally because the spatial template at $M_{500c}\lesssim10^{13}$~M$_\odot$ is not the same as in massive clusters that motivate the `universal pressure profile'. In our zoom-in simulations, this effect is not large enough to bring all of our haloes into complete agreement (i.e.\ to within $1\sigma$) with the \citet{Planck2013} observations at $5R_{500c}$. Given the observational uncertainties \citep{Greco2015} we do not draw strong conclusions from this small discrepancy at this time. If it is confirmed by future observations, it is possible that it is necessary to include line-of-sight gas outside $5R_{500c}$, which our simulations are lacking. Another possibility is that our three haloes happen to have low SZ flux and with a larger sample we would recover the mean value. Finally, this could point towards feedback being too effective at $M_{500c}\approx10^{13}$~M$_\odot$ in our current stellar feedback only simulations.

Ultimately, observations with better angular resolution are necessary to probe the effect of feedback on hot haloes around galaxies using SZ measurements. The reduction of $\widetilde{Y}_{500c}$ we find compared to the self-similar solution is stronger at lower halo masses, but already present in group-sized haloes ($M_{500c}\approx10^{13}$~M$_\odot$; $M_\mathrm{star}\approx10^{11}$~M$_\odot$) within the virial radius.

We find a steep relation between the soft X-ray luminosity, $L_X$, and galaxy or halo mass: $L_X\propto M_\mathrm{star}^{2.7}$ and $L_X\propto M_{500c}^{2.7}$ (Fig.\ \ref{fig:X_mass}). The scatter in $L_X$ between different haloes and the time-variability of $L_X$ increase significantly towards low halo masses. Our combined simulation sample below $M_{500c}=10^{12}$~M$_\odot$ shows a steep correlation between $L_X$ and SFR (Fig.\ \ref{fig:X_SFR}), becoming steeper towards lower halo or stellar mass (Fig.\ \ref{fig:X_slope}). However, the galaxies in more massive haloes show no correlation between $L_X$ and SFR. This difference arises because in the latter case, the quasi-hydrostatic, virialized halo gas dominates the X-ray emission, whereas in the former case the emission is primarily powered by star formation-driven outflows. 

The X-ray emission is highly biased towards gas with temperatures $\gtrsim10^6$~K. In low-mass haloes, gas can only reach temperatures this high if it is heated by galactic winds (directly or indirectly). For $M_{500c}\lesssim10^{11.5}$~M$_\odot$ the X-ray-weighted radial velocity is close to the halo's circular velocity, showing that indeed most of the emission is coming from outflowing gas (Fig.\ \ref{fig:X_vrad}). In high-mass haloes, the virial temperature is high enough ($\gtrsim10^6$~K) for quasi-hydrostatic halo gas, shock heated at the virial radius, to contribute to (and dominate) $L_X$, resulting in X-ray-weighted radial velocities much lower than $v_\mathrm{circ}$. 

Comparing our soft X-ray predictions to stacked observations by \citet{Anderson2015}, we find that they are consistent within the scatter, although we are limited by statistics. 
\citet{Brun2015} showed that their simulation (with lower resolution, but much better halo statistics) with only stellar feedback overpredicts $L_X$ for $M_{500c}\approx10^{13}$~M$_\odot$. This is not true in our sample of higher-resolution zoom-in simulations. This is likely because our stellar feedback implementation is more efficient at driving galactic winds in group-sized haloes. The simulation of \citet{Brun2015} including both stellar and AGN feedback also reproduces observations well. This paper shows that AGN feedback may not be necessary to explain the X-ray properties of galaxy groups. Note, however, that the SFRs of our massive galaxies lie at the high end of those observed in local early-type galaxies \citep{Davis2014} and additional feedback may be required to reproduce the full quenched galaxy population, without substantially changing the X-ray luminosities. 

\citet{Crain2010a} find that the X-ray luminosity of their simulated galaxies (with $10^{10}<M_\mathrm{star}<10^{11.7}$~M$_\odot$) is not driven by the present-day SFR, although it is still biased towards outflowing gas, because outflowing gas is denser and more metal-rich. Our simulations agree with this claim at high masses ($M_\mathrm{star}\approx10^{11}$~M$_\odot$), but show evidence that for $M_\mathrm{star}\approx10^{10}$~M$_\odot$ the X-ray emission is enhanced at high SFR. The difference could be due to the fact that the FIRE simulations better resolve the galaxy's interstellar medium and the sites of star formation, which results in a more time-variable, stochastic SFR \citep{Hopkins2014FIRE, Sparre2015} compared to lower-resolution simulations in which the high-density gas is modelled with an effective equation of state \citep[as in][]{Crain2009, Crain2010a}. 

We caution that the X-ray emission from diffuse gas we calculate from star-forming galaxies may be an underestimate. The reason for this is that most of the X-ray emission from galactic winds driven by star formation is predicted to come from small radii comparable to the size of the star-forming disc \citep[e.g.][]{Zhang2014}. This is consistent with observations of NGC 1569 \citep{Martin2002} and M82 \citep{Strickland2009}. This emission probably originates primarily from the interfaces between the volume filling supernovae heated gas and embedded cool `clouds' \citep[e.g.][]{Veilleux2005}. Properly resolving these dynamics is challenging for simulations of isolated galaxies or idealized patches of galactic disks \citep[e.g.][]{Martizzi2016}, let alone for cosmological simulations. The inclusion of magnetic fields and/or thermal conduction may also be critical for capturing the correct dynamics at the interface between the cool and hot phases \citep[e.g.][]{McCourt2015, Bruggen2016}. Thus, although we do not see any trend in our predicted X-ray emission with resolution, it is possible that this emission would increase at much higher resolution and with the inclusion of additional physics. We expect, however, that the general trends with SFR and stellar mass found here are robust, since they are caused by the change in the relative importance of star formation and hot halo gas with halo mass, which is a generic feature of galaxy formation.

Our current cosmological zoom-in simulations predict that future observations of the SZ effect with high angular resolution (e.g.\ SPTpol, ACTpol, SPT-3G, AdvACT, NIKA), will measure a reduced $\widetilde{Y}$ (compared to the self-similar value) in $<10^{13}$~M$_\odot$ haloes due to gas ejection from the halo. Future surveys of the diffuse soft X-ray emission (e.g.\ Chandra, XMM-Newton, eROSITA, ATHENA, SMART-X) around Milky Way-mass and dwarf galaxies will observe increased scatter in $L_X$ towards lower masses and a strong correlation with SFR. Combining both observational probes is a particularly strong diagnostic of the effect of stellar feedback on the gaseous haloes around galaxies, because galactic winds affect the SZ signal and soft X-ray luminosity differently. In the future we aim to extend our current simulation suite with AGN feedback to study its effect and with a larger volume cosmological simulation in order to probe SZ and X-ray properties statistically, compare in detail to the observed scatter, and study gas outside the virial radius.

\section*{Acknowledgements}

We would like to thank the referee for constructive comments, the Simons Foundation and participants of the Simons Symposion `Galactic Superwinds: Beyond Phenomenology' for inspiration for this work, and Tim Davis for helpful comments on an earlier version of the manuscript. 
FvdV also thanks Rob Crain and Ian McCarthy for useful discussions. 
EQ was supported by NASA ATP grant 12-APT12-0183, a Simons Investigator award from the Simons Foundation, and the David and Lucile Packard Foundation.
Support for PFH was provided by an Alfred P. Sloan Research Fellowship, NASA ATP Grant NNX14AH35G, and NSF Collaborative Research Grant \#1411920 and CAREER grant \#1455342. 
CAFG and ZHH were supported by NSF through grants AST-1412836 and AST-1517491, by NASA through grant NNX15AB22G, and by STScI through grants HST-AR-14293.001-A and HST-GO-14268.022-A.
DK was supported by NSF grant AST-1412153.
The simulations presented here used computational resources granted by the Extreme Science and Engineering Discovery Environment (XSEDE), which is supported by NSF grant number OCI-1053575, specifically allocations TG-AST120025 (PI Kere\v{s}), TG-AST130039 (PI Hopkins), TG-AST1140023 (PI Faucher-Gigu\`ere) as well as the Caltech computer cluster `Zwicky' (NSF MRI award \#PHY-0960291) and the Northwestern computer cluster Quest.
Some simulations were run with resources provided by the NASA High-End Computing (HEC) Program through the NASA Advanced Supercomputing (NAS) Division at Ames Research Center, proposal SMD-14-5492.

\bibliographystyle{mnras}
\bibliography{szxray}

\bsp

\appendix

\section{Radial soft X-ray profiles} \label{sec:prof}

\begin{figure}
\center
\includegraphics[scale=.52]{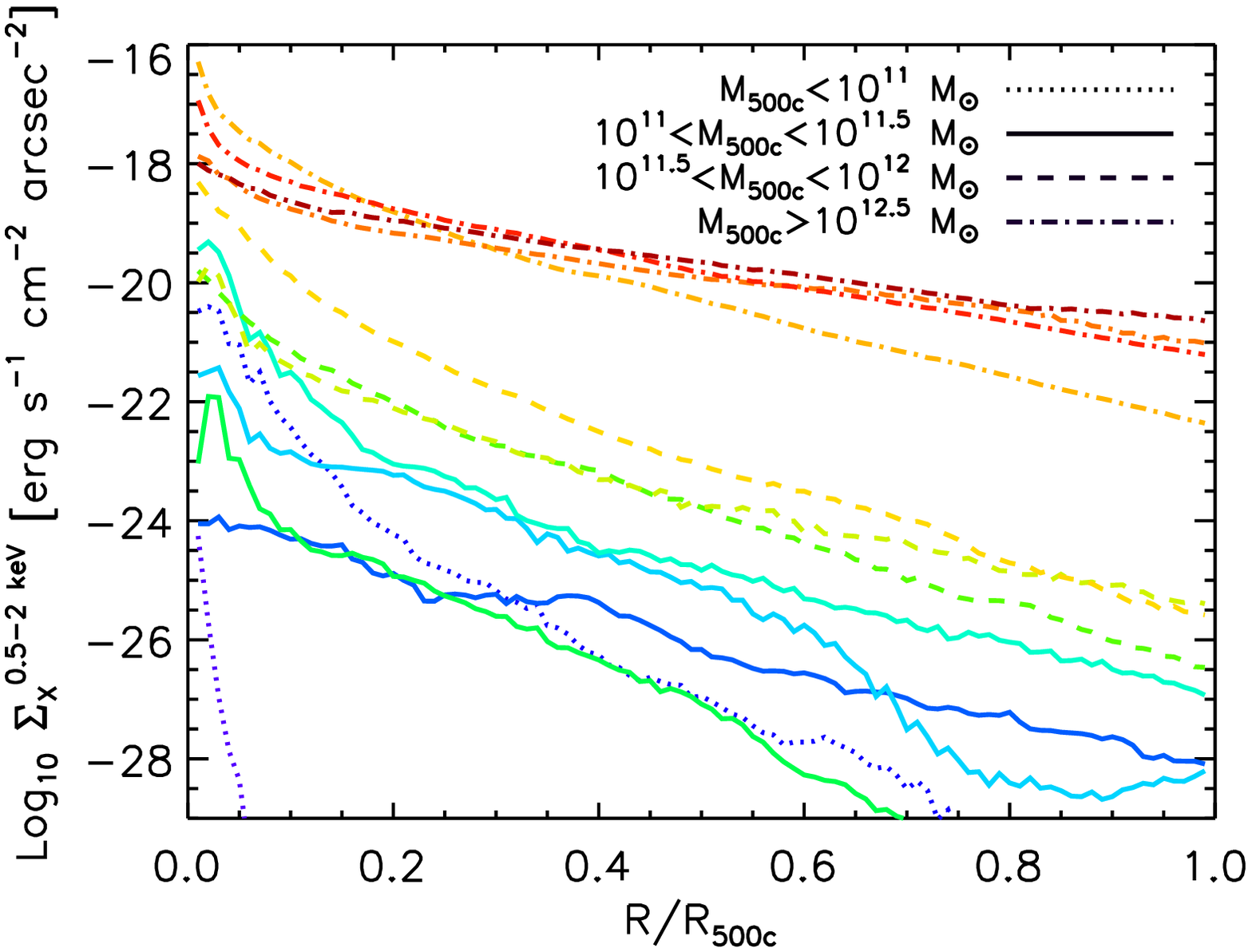}
\caption {\label{fig:X_prof} Median radially averaged soft X-ray surface brightness profile from $z=0.5-0$ for all 16 galaxy haloes considered in this paper. Colours correspond to those in Figures~\ref{fig:frac} to~\ref{fig:X_vrad}. Dotted curves correspond to low-mass haloes ($M_{500c}<10^{11}$~M$_\odot$), solid curves to $10^{11}<M_{500c}<10^{11.5}$~M$_\odot$, dashed curves to $10^{11.5}<M_{500c}<10^{12}$~M$_\odot$, and dot-dashed curves to massive haloes ($M_{500c}>10^{12.5}$~M$_\odot$). The profiles of the star-forming galaxies are fairly steep, with the emission powered by galactic winds at small radii. For massive galaxies, however, the profiles are much shallower, showing that for these objects the halo outskirts contribute a larger fraction of the total X-ray luminosity.}
\end{figure} 
This appendix shows predictions for soft X-ray surface brightness profiles that can be directly compared to (future) observations. Figure~\ref{fig:X_prof} shows the radially averaged soft X-ray surface brightness profile where we have taken the median from $z=0.5-0$. Colours show individual haloes and are identical to those in all previous Figures. Dotted curves correspond to low-mass haloes ($M_{500c}<10^{11}$~M$_\odot$), solid curves to $10^{11}<M_{500c}<10^{11.5}$~M$_\odot$, dashed curves to $10^{11.5}<M_{500c}<10^{12}$~M$_\odot$, and dot-dashed curves to massive haloes ($M_{500c}>10^{12.5}$~M$_\odot$).  A clear difference can be seen in the slope of the median profiles, which are shallower for the four group-sized haloes. For these systems, the halo outskirts contribute a significant fraction of the total X-ray luminosity, whereas for star-forming galaxies, the emission is more centrally concentrated. This again argues for different origins of the X-ray emission with hot gas in massive systems being predominantly heated by an accretion shock at the virial radius, whilst the hot gas in lower-mass, star-forming systems is heated primarily by stellar feedback in the centre.

\begin{figure}
\center
\includegraphics[scale=.52]{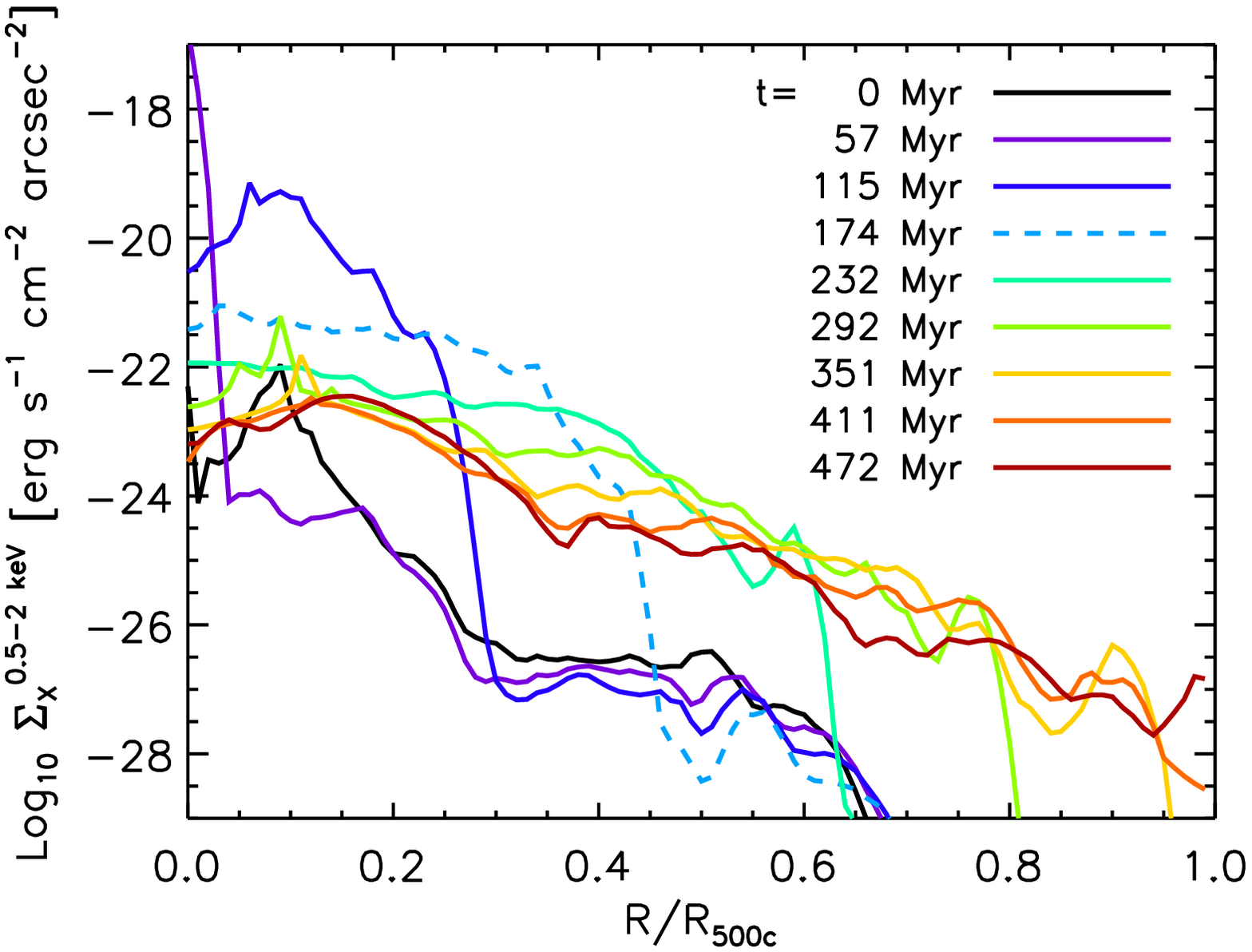}
\caption {\label{fig:X_m11} Radially averaged soft X-ray surface brightness profile of galaxy m11 at different times, with time intervals around 60~Myr as shown in the legend. The blue, dashed curve corresponds to the image in Figure~\ref{fig:img}. When a feedback event increases the total X-ray luminosity, this is visible as a sharp drop in the surface brightness profile. This drop propagates outwards as the hot gas expands. About 300~Myr after the hot wind was triggered, the profile becomes smooth again.}
\end{figure} 
Because the time variability is large for low-mass galaxies, the soft X-ray profile at any given time can look very different from the median profile shown in Figure~\ref{fig:X_prof}. To illustrate this, we show the radially averaged soft X-ray profile for galaxy m11 in Figure~\ref{fig:X_m11}. The different curves correspond to different times, with $\sim60$~Myr intervals. The blue, dashed curve shows the X-ray surface brightness correspoding to the image shown in Figure~\ref{fig:img}. For this galaxy, at $t=57$~Myr, stellar feedback increases the total X-ray luminosity in the centre, which propagates outwards as the hot gas expands. This expanding bubble is visible as a sharp drop in the surface brightness profile. After a few 100~Myr, the profile becomes smooth again. The soft X-ray profiles can change significantly on 100~Myr timescales for galaxies with $M_\mathrm{star}<10^{10}$~M$_\odot$. The variation seen in Figure~\ref{fig:X_m11} is smaller for more massive systems, but in all our simulations, a sharp drop in the X-ray surface brightness profile of a galaxy halo is strong evidence for stellar feedback powering the X-ray emission at least interior to the drop.

\label{lastpage}

\end{document}